\documentclass[12pt]{iopart}

\usepackage{graphicx}
\usepackage{epstopdf}

\usepackage[T1]{fontenc}
\usepackage{libertine}
\expandafter\let\csname equation*\endcsname\relax
\expandafter\let\csname endequation*\endcsname\relax
\usepackage{amsmath}
\usepackage{amssymb}
\usepackage{cite}

\usepackage{color}

\newcommand{\eins}{\mbox{$1 \hspace{-1.0mm} {\bf l}$}}

\begin{document}
 
\title[Time-convolutionless stochastic Schr\"odinger
equation]{Application of a time-convolutionless stochastic
Schr\"odinger equation to energy transport and thermal relaxation}

\author{R. Biele$^{1,2}$, C. Timm${^2}$ and R. D'Agosta$^{1,3}$}
\address{$^1$ETSF Scientific Development Center, Departamento de F\'isica de
Materiales, Universidad del Pa\'is Vasco, E-20018 San Sebasti\'an, Spain}
\address{$^{2}$Institute of Theoretical Physics, Technische Universit\"at
Dresden, D-01062 Dresden, Germany}
\address{$^3$IKERBASQUE, Basque Foundation for Science, E-48011, Bilbao, Spain}
\date{\today}

\begin{abstract}
Quantum stochastic methods based on effective wave functions form a framework
for investigating the generally non-Markovian dynamics of a quantum-mechanical
system coupled to a bath. They promise to be computationally superior to the
master-equation approach, which is numerically expensive for large dimensions
of the Hilbert space. Here, we numerically investigate the suitability of a
known stochastic Schr\"odinger equation that is local in time for the
description of thermal relaxation and energy transport. This stochastic
Schr\"odinger equation can be solved with a moderate numerical cost, indeed 
comparable to that of a Markovian system, but reproduces the dynamics of a system
evolving according to a general non-Markovian master equation. After verifying
that it describes thermal relaxation correctly, we apply it for the first time
to the energy transport in a spin chain. We also discuss a portable algorithm
for the generation of the coloured noise associated with the numerical solution of the non-Markovian dynamics.
\end{abstract}
%\pacs{05.40.-a, 42.50.Lc, 05.10.Gg}
\maketitle 

\section{Introduction}
\label{intro}

Stochastic methods applied to the investigation of the dynamics of physical
systems coupled to external baths have a long history dating back to Einstein
\cite{Einstein1905} and Langevin \cite{Langevin1908}. The idea behind these
approaches is that the many degrees of freedom of the bath induce random motion
in the system
\cite{Kloeden1999,Gardiner2000,Breuer2002,Razavy2006,vanKampen,Weiss2007}.
Classically, this stochastic motion is due to collisions between the particles
of the system and of the bath and can be described by a Langevin equation for
certain system variables. Quantum mechanically, the randomness is introduced by
transitions between different states of the system induced by the bath and can
be described by a stochastic Schr\"odinger equation (SSE)
\cite{vanKampen,Ghirardi1990,Diosi1997,Gaspard1999,Yu1999,Strunz2000}.
Alternatively, one can derive statistical descriptions averaging over many
realisations of the stochastic process, leading to the Fokker-Planck equation
for the distribution function of a classical system and the master equation for
the reduced density operator of a quantum system
\cite{Kloeden1999,Gardiner2000,Breuer2002,Razavy2006,vanKampen,Weiss2007},
respectively. Assuming the equivalence between the master equation for the
density matrix $\hat{\rho}(t)$ and the SSE \cite{Yu1999,vanKampen}, which might
not always hold \cite{DAgosta2008a}, the latter has sometimes been seen as a
``quick and dirty'' way to obtain the solution of the former. Indeed, the
numerical solution of the master equation scales poorly with the number of
states kept in the calculation since it is an equation of motion for a
\emph{matrix} in state space, whereas a Schr\"odinger equation is an equation
for a \emph{vector}. This strongly limits the applicability of the master
equation to complex systems. In particular, if the quantum system consists of
interacting particles, in which case the state space is the Fock space, the
scaling of the density matrix restricts the range of system that could be
tackled with numerical calculations and approximate methods are needed
\cite{Pershin2008}. Alternatively, one can establish a time-dependent density
functional theory \cite{Marques2006} of open quantum systems
\cite{Burke2005,DiVentra2007,DAgosta2008a}.

In general, the dynamics of an open system is non-Markovian, i.e., the change
of the state of the system at a certain time does not only depend on its
present state but also on its state at all previous times. Understandably, the
solution of a non-Markovian master equation \cite{Nak58,Zwa60} is difficult
because it involves the evaluation of a convolution integral which depends on
the history of the system. Therefore, one often employs a Markov approximation,
which replaces the dynamical memory kernel in this convolution integral by a
$\delta$-function in time. In doing so, however, one looses the connection with
the exact dynamics and the ability to reproduce the correct steady state,
unless one is capable of constructing effective bath-system operators that
recover the exact behaviour \cite{Wichterich2007}. It should be noted that an
exact time-convolutionless master equation can be derived
\cite{ToM76,STH77,Breuer2002,Timm2011}. However, this does not usually reduce
the numerical complexity since one needs to evaluate the generator of the
time-convolutionless master equation describing the history of the system at
each time step of the numerical integration.

In order to study the non-Markovian dynamics it would be advantageous to have a
SSE that is local in time but is nevertheless able to reproduce the dynamics
induced by a non-Markovian master equation (NMME). Such an equation has been
proposed by Strunz and coworkers: first mentioned as a byproduct in
\cite{Strunz2004}, it has been applied to the spin-boson model and compared to
non-linear SSEs in \cite{DeVega2005} and also to a more realistic two-level
model immersed in a photonic band-gap material in \cite{DeVega2005b}. Here, we
arrive at the same time-convolutionless SSE (TCLSSE) of Strunz and coworkers starting from a non-Markovian
SSE obtained by Gaspard and Nagaoka \cite{Gaspard1999}. We show how the
dynamics induced by the TCLSSE and the NMME coincide up to third order in the
coupling parameter between the system and the bath.

A promising application of the formalism is the investigation of the
bath-induced energy transport in the system. For this the system is coupled
locally at its ends to two baths kept at different temperatures. The
temperature gradient induces a thermal force leading to energy transport in the
system. Before investigating this non-equilibrium situation within the TCLSSE
approach, we will first test whether the TCLSSE is able to reproduce relaxation
dynamics correctly: In contact with a single bath at a constant temperature,
the system should approach an equilibrium state with that temperature. It can
be shown that in the non-Markovian case there exists an exact condition that
the memory kernel must satisfy for the system to reach thermal equilibrium,
i.e., $\hat \rho(t\rightarrow \infty)\propto \exp(-\beta \hat H)$
\cite{Breuer2002,Biele2011a}. \footnote{It is well known that the Hamiltonian
$\hat H$ appearing here might be different from the one describing the system
dynamics, due for example to the Stark and Lamb shifts. For simplicity, we here
assume that these effects can be neglected, since they are normally
proportional to $\lambda^4$, where $\lambda$ is the coupling parameter between
the system and the bath.} This condition is known as \emph{detailed balance}
since it relates the absorption and emission probabilities. The
detailed-balance condition is usually no longer satisfied if the Markov
approximation is made \cite{Breuer2002,Biele2011a}. Hence, the history
dependence of the equation of motion is an essential ingredient for thermal
relaxation. This begs the question of whether a TCLSSE is also able to
correctly describe thermal relaxation dynamics. To answer this question, we
study the relaxation of a simple three-level system employing the TCLSSE and
compare its dynamics to the one obtained from the NMME. This rather pedagogical
study verifies the applicability of the TCLSSE to thermal transport, where we
study for the first time the non-Markovian time evolution of the energy current
induced by a thermal gradient.

The numerical solution of the TCLSSE requires the generation of complex
coloured noise to mimic the correlation functions of the
non-Markovian baths \cite{Luczka2005}. Here we introduce a portable and fast
algorithm to generate any coloured noise whose power spectrum is a positive
function. The algorithm relies on the ability of performing a fast Fourier
transform and is therefore easily optimised. Other algorithms have been
presented in the past to generate real coloured noise
\cite{Rice1944,Billah1990,Barrat2011}. In section \ref{noisegeneration}, we
compare our algorithm to some of them and show that it performs better than
these routines while having a broader range of applicability.

\section{Method}
\subsection{A time-convolutionless stochastic Schr\"odinger equation}

Our starting point is a standard second-order NMME
\cite{Nak58,Zwa60,Gardiner2000,Breuer2002,Weiss2007}. The coupling between the
system and the bath is taken to be bilinear,
\begin{equation}
\hat H_\mathrm{int}=\lambda\sum_a \hat S_a \otimes \hat B_a,
\label{H_int}
\end{equation}
in the operators $\hat S_a$ and $\hat B_a$ from the system and the bath,
respectively. If any operator of the system commutes
or anticommutes with any operator of the bath, one
can always expand any coupling operator in this form.

In the following we assume that the bath and the system do not exchange
\emph{fermions}, i.e., $\hat S_a$ and $\hat B_a$ commute with each other. We
further restrict ourselves to the case that $\hat S_a$ and $\hat B_a$ are
Hermitian operators; the extension to the more general case where only $\hat
H_\mathrm{int}$ is Hermitian is straightforward. Under the assumptions of weak
system-bath interaction, factorisation of the full density operator at the
initial time $t=0$ and vanishing averages of bath operators to first order, the
equation of motion for the reduced density operator $\hat\rho$ of the system is
given by \cite{Gardiner2000,Breuer2002,vanKampen,Weiss2007}
\begin{equation}
\frac{d\hat \rho(t)}{dt}= -i\big[\hat H,\hat \rho(t)\big]
  +\lambda ^2  \sum_{a}\, [\hat S_a, \hat M^{\dagger}_a (t)-\hat M_a (t)] ,
\label{nmme}
\end{equation}
 up to second order in the coupling parameter $\lambda$.
We have set $\hbar=1$ and defined
\begin{equation}
\hat M_a (t) \equiv \sum_b \int_0^{t} d \tau\,   c_{a b}(t,\tau)\,
  e^{-i\hat H (t-\tau)}\, \hat S_b\, \hat \rho(\tau)\, e^{i \hat H(t-\tau)}.
\label{nmme_M}
\end{equation}
In this NMME, $\hat H$ is the Hamiltonian of the system and the correlation
kernel is given by
\begin{equation}
c_{a b}(t,\tau) \equiv \mbox{Tr}_B[\hat
  \rho_B^{\mathrm{eq}}\, \hat B_{a}(t)\, \hat B_{ b}(\tau)] ,
\label{cBB}
\end{equation}
where the trace is over the bath degrees of freedom, $\hat B_a(t) \equiv
e^{i\hat H_B t} \hat B_a e^{-i\hat H_B t}$ and $\hat H_B$ is the Hamiltonian
of the bath. Here, $\hat \rho_B^{\mathrm{eq}}$ is the statistical operator of
the bath. If $\hat \rho_B^{\mathrm{eq}}$ describes a single bath in thermal
equilibrium, $\hat \rho_B^{\mathrm{eq}}\propto \exp(-\beta \hat H_B)$, where
$\beta$ is the inverse temperature, the system should relax towards thermal
equilibrium, $\hat \rho(t\rightarrow \infty)\propto \exp(-\beta \hat H)$, with
the same temperature as the bath. The property that if a steady state exists,
it coincides with the state of thermal equilibrium must be encoded in the
correlation kernel $c_{a b}(t,\tau)$. Indeed, one can show that the system
relaxes towards thermal equilibrium if $c_{ab}(t,\tau)=c_{ab}(t-\tau)$ and the
power spectrum $C_{ab}(\omega)\equiv\int_{-\infty}^{+\infty} dt\, c_{a
b}(t)\,e^{-i\omega t}$ satisfies the detailed-balance condition
\cite{Breuer2002,Biele2011a}
\begin{equation}
C_{ab}(-\omega) = e^{\beta \omega}\, C_{b a}(\omega).
\label{detailed_balance}
\end{equation}

Gaspard and Nagaoka \cite{Gaspard1999} have shown that the
dynamics introduced by the NMME can be obtained not only by a
numerical integration
of (\ref{nmme}) but also by the solution of a
SSE for a state $|\Psi(t)\rangle$,
\begin{eqnarray}
i\frac{d}{dt}
 |\Psi(t)\rangle &=& \hat H |\Psi(t)\rangle + \lambda
\sum_a \gamma_a(t)\, \hat S_a|\Psi(t)\rangle  \nonumber \\
&&{} -i\, \lambda^2\sum_{a, b} \hat S_{a} \int_0^t dt'\,c_{a b}(t')\,
  e^{-i\hat H t'}\hat S_b |\Psi(t-t')\rangle.
\label{non-markovian}
\end{eqnarray}
In this non-Markovian SSE (NMSSE), the complex noises
$\gamma_{a}(t)$ have the properties
\begin{equation}
\overline{\gamma_a(t)}=0,\quad
\overline{\gamma_{a}(t)\gamma_{ b}(t')}=0,\quad
\overline{\gamma_{a}^*(t)\gamma_{ b}(t')}=c_{a b}(t-t')
\label{color_noise}
\end{equation}
and one can obtain the dynamics of the open quantum system by
taking the average over
realisations of the stochastic process, indicated by the overline. In
particular, the reduced density operator is obtained as $\hat \rho(t) =
\overline{|\Psi(t)\rangle \langle \Psi(t)|}$. However, any attempt to solve the
NMSSE (\ref{non-markovian}) requires a large numerical effort due to the time
integral, which needs to be evaluated at every time step and for every
realization. This begs the question of whether there exists a
simpler SSE that reproduces on average the dynamics induced by the NMME.
This is the case, as Strunz and Yu have shown
\cite{Strunz2004}.

Indeed, the TCLSSE
\begin{equation}
i\frac{d}{dt}
 |\Psi(t)\rangle=\bigg(\hat H + \lambda\sum_a
\gamma_{a}(t)\,\hat S_a - i\,\lambda^2\, \hat T(t)\bigg)|\Psi(t)\rangle
\label{non-markovian2}
\end{equation}
with
\begin{equation}
\hat T(t) \equiv \sum_{a, b}\hat S_a\int_0^t dt'\,c_{a b}(t')\,e^{-i\hat H
  t'}\hat S_b\, e^{i\hat H t'}
\end{equation}
reproduces on average the dynamics induced by the NMME (\ref{nmme}) up to
third order in $\lambda$ \cite{Strunz2004}.
To prove this, we write (\ref{non-markovian2}) in the
interaction picture, $|\Psi_I(t)\rangle= e^{i \hat H t}\,|\Psi(t)\rangle$
and $\hat S_a(t)=e^{i \hat H t} \hat S_a\, e^{-i \hat H t}$ and expand the
time-evolution operator up to second order in $\lambda$,
\begin{eqnarray}
|\Psi_I(t)\rangle &\cong& \left[ \eins -i
  \lambda\sum_{a}  \int_0^t d t_1 \,\gamma_{a}(t_1)\, \hat S_{a}(t_1)
  \right.\nonumber \\
&&-\lambda^2\sum_{a,b} \int_0^t d t_1 \int_0^{t_1} d
  t_2\, c_{a b}(t_2)\, \hat S_{a}(t_1)\, \hat S_{b}(t_1-t_2) \nonumber \\
&&\left.-\lambda^2 \sum_{a,b} \int_0^t d t_1  \int_0^{t_1} d
  t_2\, \gamma_{a}(t_1)\, \hat S_{a}(t_1)\, \gamma_{b}(t_2)\, \hat S_b
  (t_2)\right]
  |\Psi_I(0)\rangle\nonumber\\
&&+\mathcal{O}(\lambda^3).
\end{eqnarray}
This expansion is inserted into the expression
for the reduced density operator
$\hat \rho_I(t) =\overline{|\Psi_I(t)\rangle \langle \Psi_I(t)|}$.
By performing the average, using the properties given in
(\ref{color_noise})
and the identity $c_{ab}(\tau,t)=c^{\ast}_{b a}(t,\tau)$,
and differentiating with respect to $t$, we arrive at
\begin{eqnarray}
\frac{d}{dt}
\hat \rho_I (t) &=&  \lambda^2 \sum_{a,b} \int_0^{t}
  d \tau\, \big[ c_{a b}(t,\tau)\, \hat S_b(\tau)\, \hat{\rho}_I(0)\,
  \hat S_a(t) \nonumber \\
&& {}- c_{a b}(t,\tau)\, \hat S_a(t)\,  \hat S_b(\tau)\, \hat \rho_I(0)
  \nonumber \\
&& {}+ c^{\ast}_{a b}(t,\tau)\, \hat S_a(t)\, \hat{\rho}_I(0)\, \hat S_b(\tau)
  \nonumber \\
&& {}- c^{\ast}_{a b}(t,\tau)\, \hat{\rho}_I(0)\, \hat S_b(\tau)\, \hat S_a(t)
  \big] + \mathcal{O}(\lambda^4).
\label{derivation_step}
\end{eqnarray}
Note that the averages of the terms in $\lambda^3$ vanish.
Furthermore, replacing $\rho_I(0)$ by $\rho_I(\tau)+\mathcal{O}(\lambda^2)$
does not change the equation up to terms of order
$\lambda^3$. Finally, by returning to the Schr\"odinger picture we arrive at
the NMME (\ref{nmme}) up to terms of order $\lambda^3$, i.e.,
higher than the order up to which these equations are valid
anyway. Indeed,
the NMME and the SSE are usually derived as a second-order expansion in
the coupling parameter $\lambda$.
This is remarkable since one might expect a more
complex time-non-local SSE to be required for
reproducing the dynamics of the NMME (\ref{nmme}).
Still, the TCLSSE is local in time, i.e., the operator
$\hat T(t)$ does not depend on the state of the system at
previous times and can thus be calculated once
before the numerical integration and be used for each
realisation of the stochastic process. Hence, the numerical cost of
solving each realisation of the TCLSSE is comparable to that of a
Markovian SSE \cite{Gaspard1999,vanKampen}. 

We note that at the same level of approximation,
$\lambda^3$, we can derive a time-convolutionless
master equation instead of the non-local equation (\ref{nmme}). Indeed, in
(\ref{derivation_step}) we could replace $\rho_I(0)$ by
$\rho_I(t)+\mathcal{O}(\lambda^2)$, arriving at
\begin{eqnarray}
\frac{d}{dt}
\hat \rho_I (t) &=&  \lambda^2 \sum_{a,b} \int_0^{t}
  d \tau\, \big[ c_{a b}(t,\tau)\, \hat S_b(\tau)\, \hat{\rho}_I(t)\,
  \hat S_a(t) \nonumber \\
&& {}- c_{a b}(t,\tau)\, \hat S_a(t)\, \hat S_b(\tau)\, \hat \rho_I(t)
  \nonumber \\
&& {}+ c^{\ast}_{a b}(t,\tau)\, \hat S_a(t)\, \hat{\rho}_I(t)\, \hat S_b(\tau)
  \nonumber \\
&& {}- c^{\ast}_{a b}(t,\tau)\, \hat{\rho}_I(t)\, \hat S_b(\tau)\, \hat S_a(t)
  \big] + \mathcal{O}(\lambda^4).
\label{tclme}
\end{eqnarray}
However, since in general we expect the density matrix and the operators $\hat
S_a$ not to commute, the integral over $\tau$ still contains
the density matrix in a complicated manner.
From a numerical point of view, the solution of this equation is
therefore not simpler than that of (\ref{nmme}). The equivalence of (\ref{nmme})
and (\ref{tclme}) is a generalization of the result
that a time-convolutionless Pauli master equation, i.e., a
master equation for the diagonal components of the density matrix only,
can be proven to be equivalent to a Nakajima-Zwanzig-Markov
Pauli master equation to second order in 
$\lambda$ \cite{KGL10,Timm2011}.

\subsection{Generation of coloured noise}
\label{noisegeneration}

The TCLSSE requires the generation of coloured noise and thus the
method will only be practicable if an
efficient algorithm for the generation of this noise is available.
Such an algorithm indeed exists, as we show below, where we extend
an algorithm presented by Rice \cite{Rice1944,Billah1990} to the
complex noise required here. We consider
only a single bath operator; the generalisation to several bath
operators is straightforward. 

Some of the existing algorithms for the generation of coloured noise rely on
the numerical solution of a stochastic differential equation that has to
produce noise with the given target correlation function $c(t)$
\cite{Luczka2005,Mannella1992}. However, such an equation is a piece of
information that
is rarely available, since even the analytic expression for
$c(t)$ may not be
known. Except for a few simple models, it is more common to have access to the
power spectrum $C(\omega)=\int_{-\infty}^{+\infty} dt\, c(t)\,e^{-i\omega t}$.
Indeed, $C(\omega)$ is connected to
the quantum transitions in the bath.

On the other hand, the algorithm presented in
\cite{Barrat2011} does not require the knowledge of a
stochastic differential equation. However, besides the power spectrum
$C(\omega)$ is does require
the inverse Fourier transform of its square root. This quantity is then
convoluted with a white noise to generate the target real coloured noise. We
will introduce an algorithm that directly uses $\sqrt{C(\omega)}$ as input,
thereby reducing the numerical cost compared to the algorithm of
\cite{Barrat2011} and that generates a complex coloured noise with the
properties given
in (\ref{color_noise}). Indeed, one can easily prove that
the noise $\gamma(t)$ can be generated by
\begin{equation}
\gamma(t)=\int_{-\infty}^{+\infty} \frac{d\omega}{\sqrt{2\pi}}\,
  \sqrt{C(\omega)}\: x(\omega)\, e^{i\omega t},
\label{generate_color_noise}
\end{equation}
where $x(\omega)$ is a white-noise process in the frequency domain
satisfying
\begin{equation}
\overline{x(\omega)}=0,\quad
\overline{x(\omega)x(\omega')}=0,\quad
\overline{x^*(\omega)x(\omega')}=\delta(\omega-\omega').
\label{white_noise}
\end{equation}
By substituting (\ref{generate_color_noise}) into $\overline{\gamma^*(t)\gamma(t')}$ and using (\ref{white_noise}), we immediately arrive at the third relation of (\ref{color_noise}).
The other relations are proven in a similar way. From a numerical point of view,
the generation of this coloured noise requires the calculation of the Fourier
transform in (\ref{generate_color_noise}).
We discretize the frequency domain with the uniform step $\Delta\omega$, and generate $x(\omega)$ by using two independent gaussian random noises with zero mean and unitary variance, $N(0,1)$, therefore $x(\omega) = (N (0, 1) + iN (0, 1))/\sqrt{2\Delta\omega}$. A similar algorithm restricted to real coloured noise has been proposed in the past \cite{Rice1944,Billah1990}.

In order to compare our algorithm with the two from \cite{Barrat2011} and
\cite{Rice1944,Billah1990},
we choose a test function for which we know $c(t)$ and $C(\omega)$
analytically, namely $c(t)=(2\pi\sigma^2)^{-1/4} e^{-t^2/2\sigma^2}$ and
$C(\omega)=(2\pi\sigma^2)^{1/4}\,e^{-\omega^2\sigma^2/2}$. We fix $\sigma=1$ as
our unit of time and choose the interval $t\in[-25,25]$ for the numerical
Fourier transform. To quantify the agreement between the target $c(t)$ and
the noise generated by the three algorithms, we use the statistical variance
$\delta_c=\int_{-\infty}^{+\infty} dt\,|c(t)-\overline{\gamma(0)\gamma^*(t)}
|^2/\int_{-\infty}^{+\infty} dt\,|c(t)|^2$. In principle, these algorithms
produce an exact representation of the target correlation function.
Discrepancies arise from the finite mesh on which the Fourier
transform is evaluated, the finite number of independent realisations
of the noise that we generate
and the limitations of the white noise generation.
\begin{figure}[ht!]
\includegraphics[width=8cm]{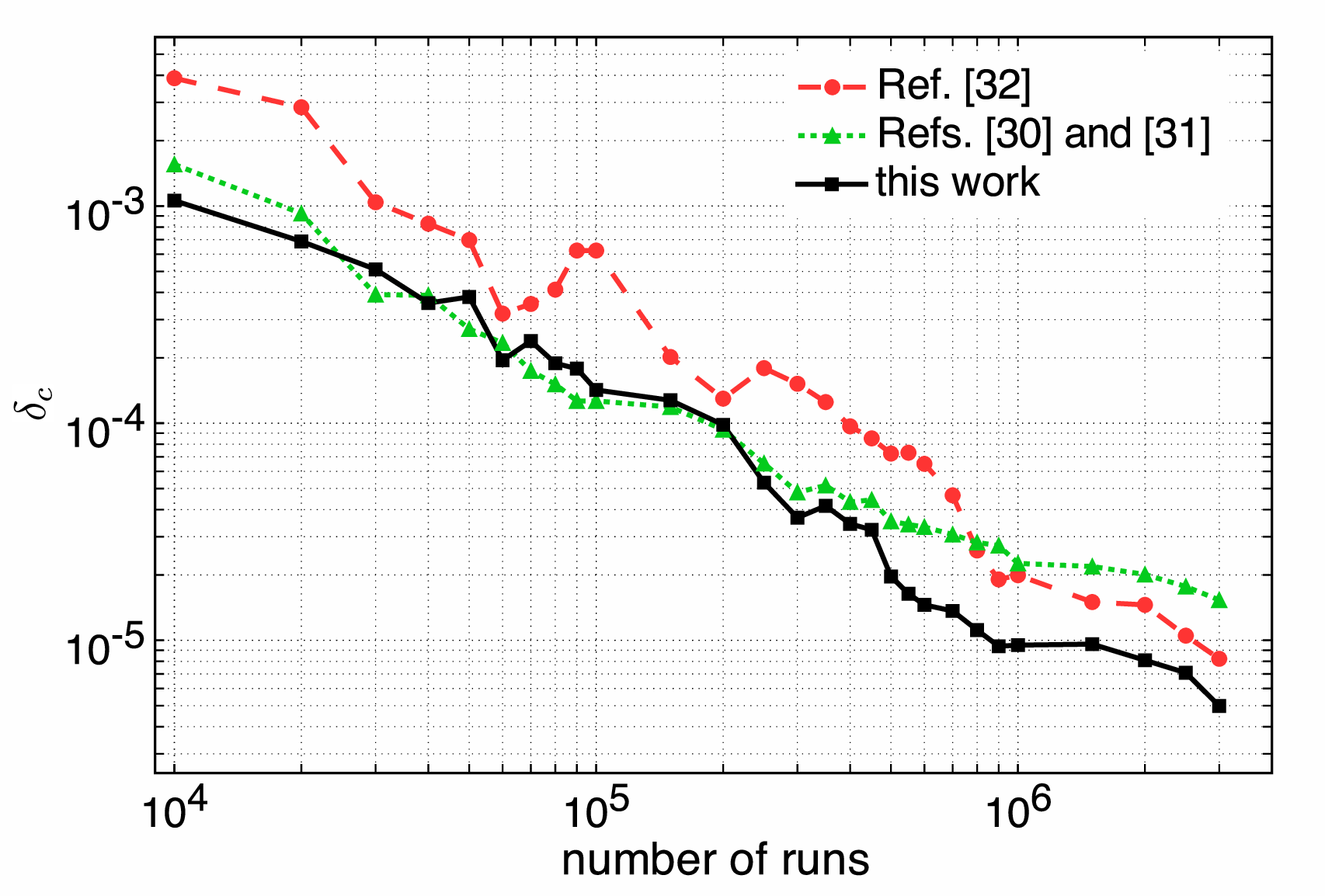}
\caption{Statistical variance $\delta_c$ versus the number of independent
realisations of the coloured noise, calculated using a
16384 point mesh in time and frequency. The red (dashed) line
represents the optimised version of the 
algorithm presented in \cite{Barrat2011}, the green (dotted) line
the algorithm proposed by Rice \cite{Rice1944,Billah1990} and the
black (solid) line the results obtained from (\ref{generate_color_noise}).}
\label{deltac}
\end{figure}

In figure \ref{deltac}, we report the variance $\delta_c$ as a function of the
number of independent realisations of the noise. We see that the algorithm
(\ref{generate_color_noise}) performs better for a large number of runs (at
least $2\times10^5$), while being close to the other two for a small number of
runs. The algorithm proposed in \cite{Barrat2011} suffers from the need of
performing a double Fourier transform, although for a large number of runs its
performance improves consistently. On the other hand, we can consider the total
computation time to generate a given number of realisations. Taking the time
needed by algorithm (\ref{generate_color_noise}) as a reference, the algorithm
of \cite{Rice1944, Billah1990} is about 7\% slower and the algorithm of
\cite{Barrat2011} is about 50\% slower. However, we stress that the main
advantage of the algorithm (\ref{generate_color_noise}) does not lie in the
moderate numerical improvement but in the simplification it brings about by
only requiring the power spectrum as input.

\begin{figure}[ht!]
\includegraphics[width=8.0cm]{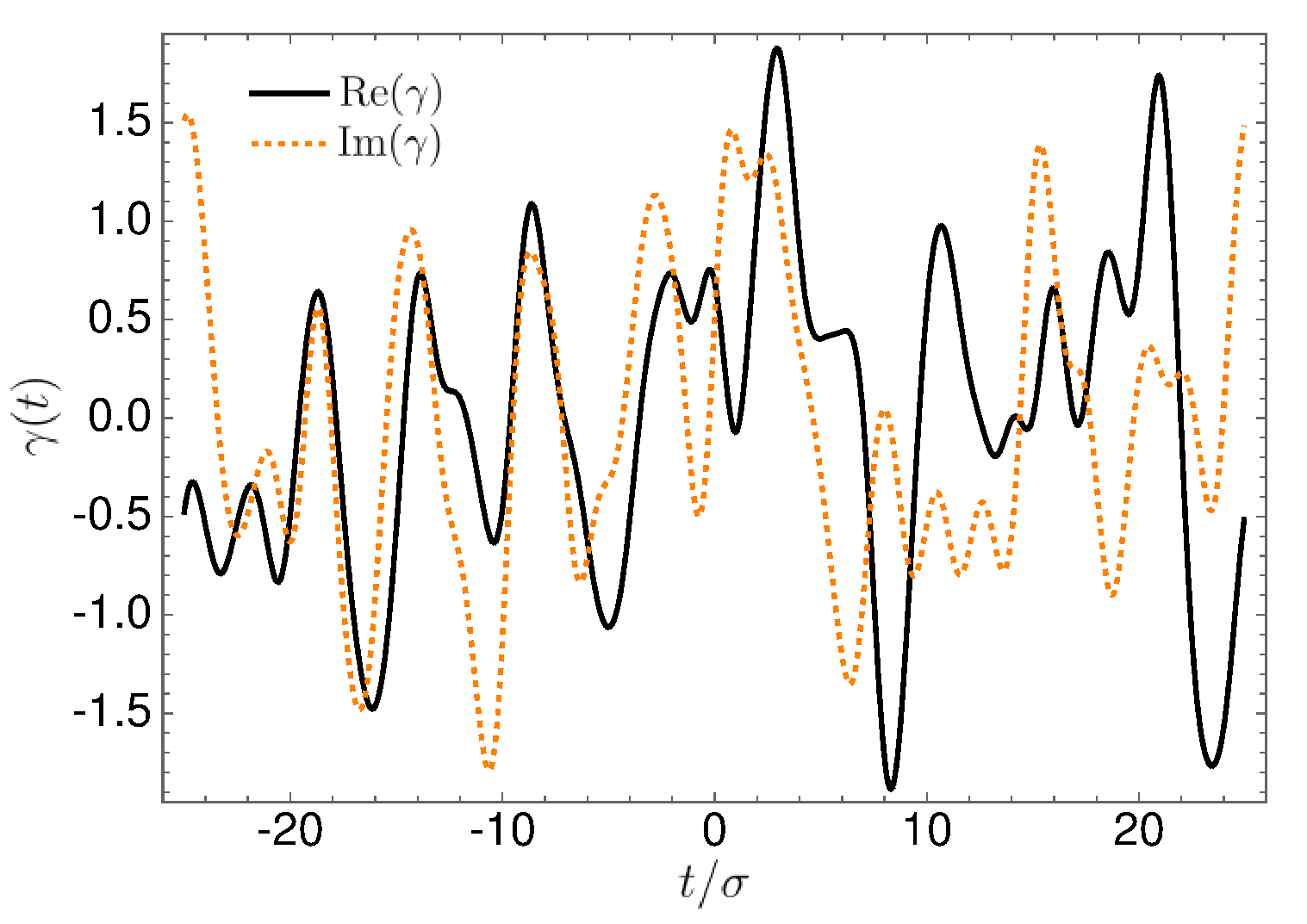}
\caption{The real (black, continuous line) and imaginary (orange, dotted line)
parts of a single realisation of the coloured noise,
(\ref{generate_color_noise}),
for a mesh in time of 16384 points. The function $\gamma(t)$ appears smooth as a
function of time due to the fact that time enters in
$\gamma(t)$ via the oscillating term in the right hand side of (\ref{generate_color_noise}).
Similar behaviours are obtained with the other two algorithms.}
\label{single_run}
\end{figure}

For illustration, we show in figure \ref{single_run} a single realisation of
the noise (\ref{generate_color_noise}) with 16834 mesh points. Notice that due
to the use of the fast Fourier transform, the noise is periodic over the
simulation time.

\subsection{Thermal relaxation of a three-level system}
Here, we test whether the TCLSSE is capable to describe thermal relaxation dynamics correctly when connected to a single bath. This will then allow us to study energy transport within the TCLSSE approach in section \ref{application}.

We consider the coupling of an electronic
system to the electromagnetic field in a three-dimensional cavity. In the
dipole approximation, one can derive from first principles the power spectrum
for this system (we set the speed of light to unity),
\begin{equation}
C_{\mathrm{cav}}(\omega)=\dfrac{|\omega|^3}{\pi \epsilon_0}
  \big[n_B(\beta|\omega|)+\theta(-\omega)\big]\quad
  \mbox{for}\quad |\omega|<\omega_{c},
\label{cavity-power-spectrum}
\end{equation}
where $n_B(\beta\omega)\equiv 1/(e^{\beta \omega}-1)$ is the
Bose-Einstein distribution function, $\theta(\omega)$
is the Heaviside step function and $\omega_c$ is a cutoff frequency 
determined by the dimensions of the system. This cutoff is
necessitated by the assumption made in the dipole approximation that the
electromagnetic field is uniform in the region of space occupied
by the system. The
derivation of this power spectrum can be found
in the appendix. For $|\omega|>\omega_c$, the power spectrum is set to vanish.
Note that increasing $\omega_c$ does not change the relaxation
dynamics as long as $\omega_c$ is larger than the energy differences in the
system and hence does not exclude any transitions. One can show that the
detailed-balance condition (\ref{detailed_balance}) is satisfied by this power
spectrum. Since for this model system the correlation function $c(t)$ is not
given in analytical form, we will use (\ref{generate_color_noise}) to generate the noise.

\begin{figure}[ht!]
\includegraphics[width=8cm]{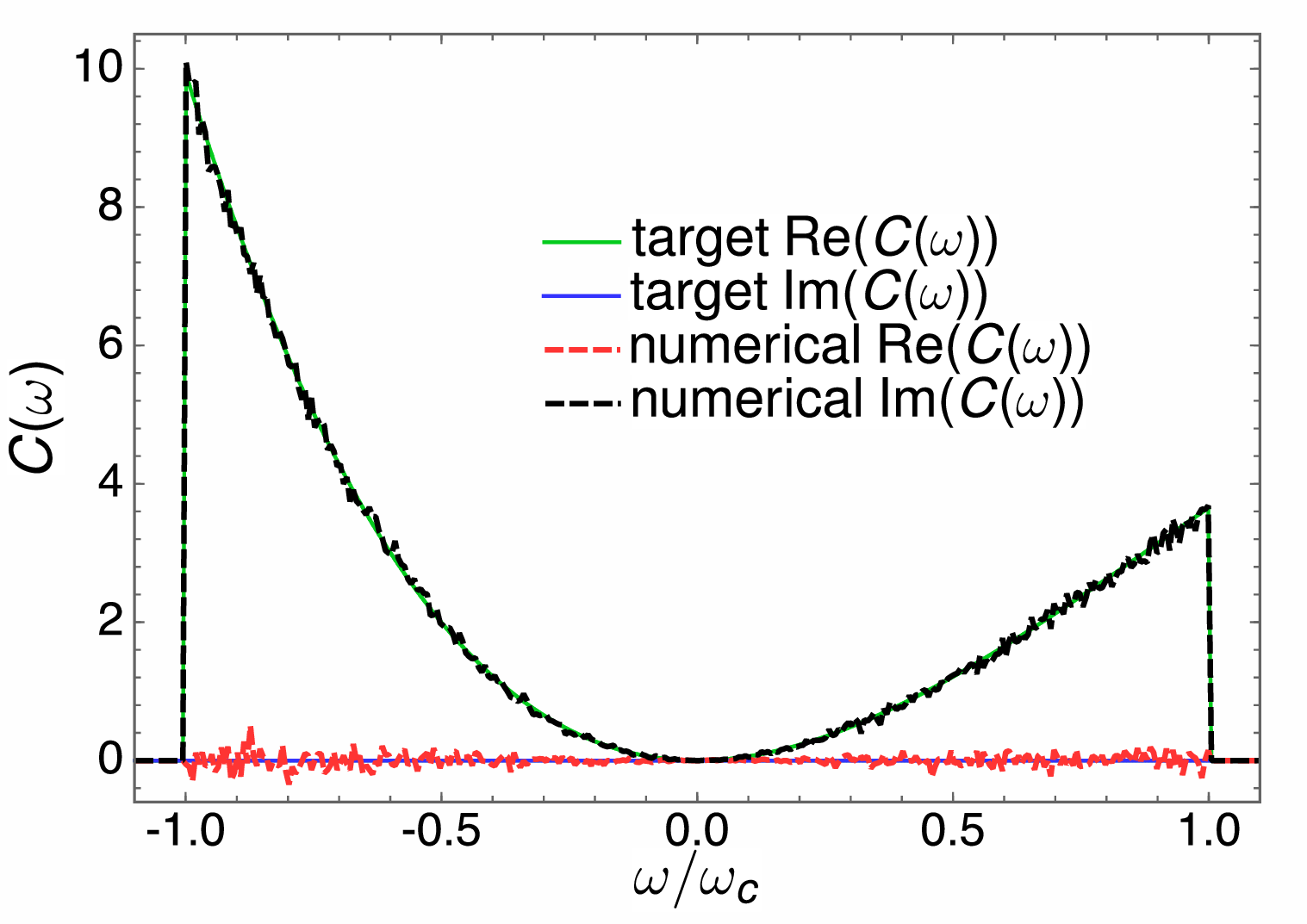}
\caption{Comparison between the target, (\ref{cavity-power-spectrum}), (solid lines) and
the Fourier transform of the correlation function obtained from (\ref{generate_color_noise}) by averaging over $90\,000$ realisations of the
noise (dashed lines).
}
\label{cofomega}
\end{figure}

In order to quantify the agreement between the noise generated by
(\ref{generate_color_noise}) and the power spectrum 
(\ref{cavity-power-spectrum}), we have performed a Fourier
transform of the time-domain signal and compared it
to our target. Figure \ref{cofomega} shows that the agreement is
excellent.

For the electronic system we consider a three-site spinless
tight-binding chain described by the Hamiltonian
\begin{equation}
\hat H =-T\,\big(\hat c^{\dagger}_1 \hat c_2+\hat c^{\dagger}_2\hat c_1 +
  \hat c^{\dagger}_2\hat c_3 + \hat c^{\dagger}_3 \hat c_2 \big),
\label{tight-binding}
\end{equation}
where the operator $\hat c^{\dagger}_i$ creates an electron at
site $i$, and assume a single electron to be present.
This system is coupled to the
electromagnetic field inside the cavity by the operator
\begin{equation}
\hat S = - q \sum_{i,j} \vec{u} \cdot \langle W_i | \vec{r} | W_j
  \rangle\, \hat c_i^{\dagger} \hat c_j ,
\label{coupling}
\end{equation}
where $q$ is the charge of the electron and $|W_i\rangle$ is the
single-particle state localized at site $i$. For simplicity, we assume that
each relevant mode of the cavity has the same polarization direction $\vec{u}$,
parallel to the tight-binding chain. Note that the form of this operator should
be immaterial for the establishment of thermal equilibrium, which is only
determined by the power spectrum. Indeed, we can check whether the
detailed-balance condition is necessary for the system to reach thermal
equilibrium. To that end, we use the operator in (\ref{coupling}) within a Markov approximation for the correlation function, $c(t)\propto \delta(t)$. We have found that a steady state is approached that does not correspond to thermal equilibrium \cite{Biele2011a}.

\begin{figure}[ht!]
\includegraphics[width=8cm]{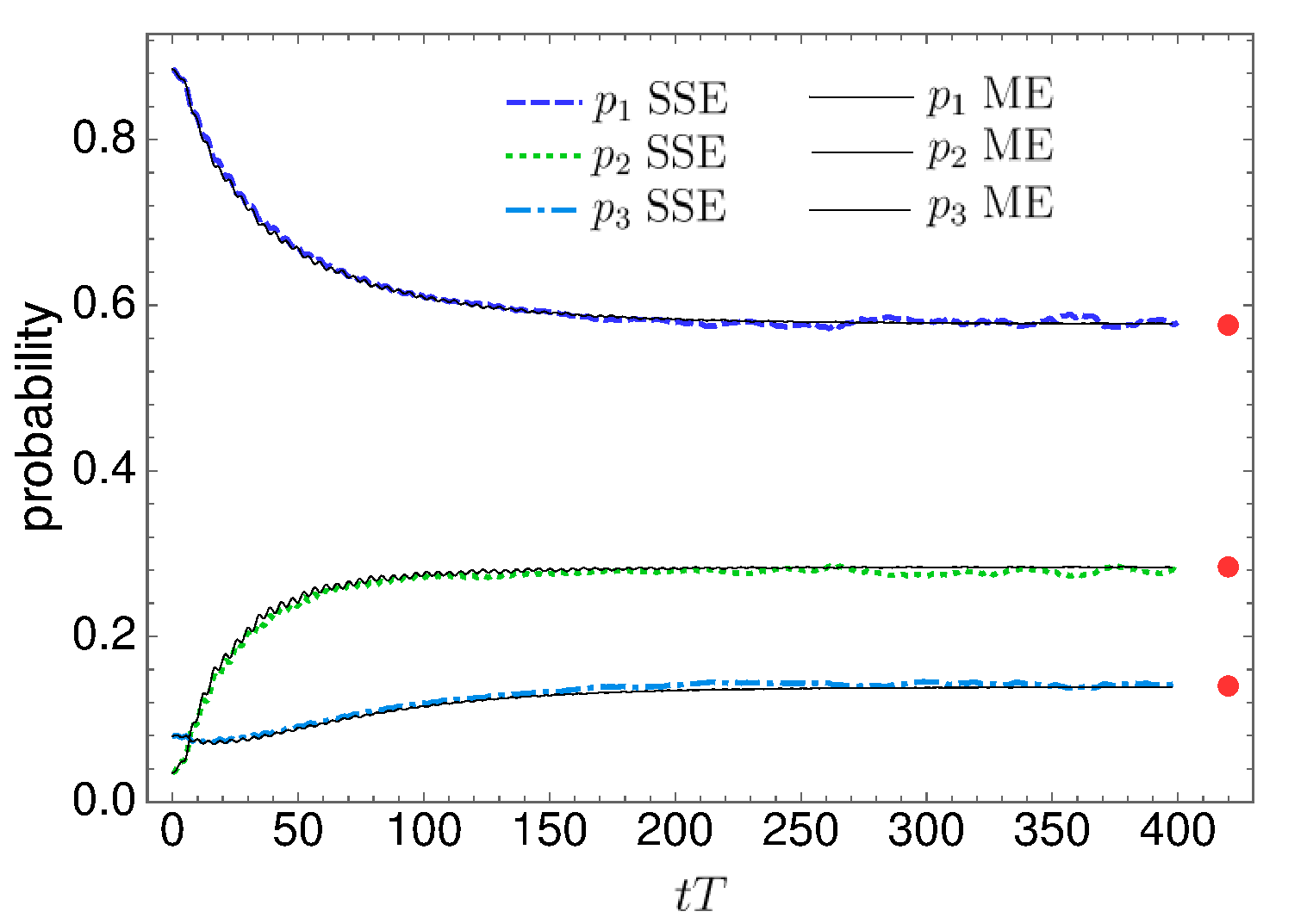}
\caption{Dynamics of the occupation probabilities $p_1$, $p_2$,
$p_3$ of
the eigenstates of the Hamiltonian (\ref{tight-binding})
in the one-electron sector calculated from
the evolution of the TCLSSE (dashed lines) and the NMME
(solid lines) with the power spectrum given by (\ref{cavity-power-spectrum}).
The eigenstates are labeled such that the eigenenergies satisfy
$\epsilon_1\leq\epsilon_2\leq\epsilon_3$. The red dots
represent the thermal-equilibrium probabilities calculated from (\ref{eq_prob}). The time $t$ is measured
in terms of the inverse of the energy constant $T$.}
\label{probability3x3}
\end{figure}

In figure \ref{probability3x3} we show the occupation
probabilities of the three eigenstates of the Hamiltonian
in the one-electron sector as a function of time calculated using
the TCLSSE (dashed lines) and the NMME (solid lines), respectively. For the
TCLSSE, the results have been obtained by
averaging over $90\,000$ independent realisations of the noise.
We have used the parameters $\beta=1$, $\omega_c=1$, $T=1$,
$\epsilon_0=1$ and $\lambda=0.1$ and
we have employed the Euler algorithm \cite{Press1992,Kloeden1997}
with time step $\Delta t=0.005$ to numerically solve the equations.
 As the establishment of thermal equilibrium is independent
of the choice of the initial state, we have chosen an arbitrary
pure state,
$| \Psi (0)\rangle = 0.94\, |1\rangle +0.2\,|2\rangle+0.28\,|3\rangle$,
where $|i\rangle$ represents the \textit{i}-th
eigenstate of the Hamiltonian, where the eigenenergies satisfy
$\epsilon_1\leq\epsilon_2\leq\epsilon_3$.

The dynamics induced by the NMME and the TCLSSE are in good
agreement: The small discrepancies in the numerical solutions are due to the
finite number of realisations we have used; the solution of the TCLSSE still
contains some noise, as expected. For long
times, both formalisms converge to the thermal-equilibrium probabilities
\begin{equation}
p_i = \frac{e^{-\beta\epsilon_i}}
  {e^{-\beta\epsilon_1} + e^{-\beta\epsilon_2} + e^{-\beta\epsilon_3}}.
\label{eq_prob}
\end{equation}
 If we were only interested in the
long-time limit, we could have averaged over all times after some
equilibration time $t_\mathrm{min}$ to obtain better statistics,
using the ergodic theorem to replace the average over many realisations
by an average over time of a single realisation.

\section{Application to energy transport}
\label{application}
To show that the TCLSSE can be used to investigate energy
transport in open quantum systems, we consider a spin chain in contact with two
baths at different
temperatures. The baths are locally connected to the terminal
spins of the
chain \cite{Wichterich2007,Monasterio2007}. Energy is transferred between the high-temperature bath, via the spin
chain, to the low-temperature bath. Here we assume the baths
to be represented by an ensemble of harmonic oscillators with a
continuous spectrum. In the long-time regime, we expect the appearance of a
steady state of constant energy flow.

The total Hamiltonian of a spin-$1/2$ chain coupled to two
baths \textit{L} and \textit{R} reads
\begin{equation}
\hat H_T = \hat H_S + \sum_{i=L,R} \big(\hat H^{(i)}_B + \hat H_{SB}^{(i)} 
\big),
\end{equation}
where the system Hamiltonian is given by
\begin{equation}
\hat H_S = \frac{\Omega}{2} \sum_{\mu=1}^n \sigma_z^{(\mu)} +
\Gamma \sum_{\mu=1}^{n-1} \vec{\sigma}^{(\mu)}\cdot \vec{\sigma}^{(\mu+1)},
\label{eq_systemH}
\end{equation}
with $\vec{\sigma}=(\sigma_x,\sigma_y,\sigma_z)$ and the index $\mu$
indicating the spin site. The Pauli matrices are given by
\begin{equation}
\sigma_x= \begin{pmatrix} 0 & 1 \\ 1 & 0 \end{pmatrix},\quad
\sigma_y= \begin{pmatrix} 0 & -i \\ i & 0 \end{pmatrix},\quad
\sigma_z= \begin{pmatrix} 1 & 0 \\ 0 & -1 \end{pmatrix}.
\end{equation}
Hence, the spin operators for the \textit{n}-site chain are
\begin{equation}
\begin{array}{ccccccccc}
& 1 & & 2& &\mu& &n \\
\sigma^{(\mu)}_{j} =& \eins & \otimes & \eins &
\otimes\cdots\otimes &\sigma_j&
\otimes\cdots\otimes &\eins.
\end{array}
\end{equation}
In (\ref{eq_systemH}), $\Omega$ is the energy associated with a
uniform
magnetic field aligned along the $z$ direction and $\Gamma$ is the spin-spin
Heisenberg interaction.

The baths are coupled to the spins at the
ends of the chain,
\begin{equation}
\hat H^{(i)}_{SB} = \lambda\, \hat S^{(i)} \otimes \hat B^{(i)}
= \lambda\, \sigma^{(i)}_x \otimes \hat B^{(i)},
\end{equation}
where $\sigma_x^{(i=L)}=\sigma_x^{(1)}$, $\sigma_x^{(i=R)}=\sigma_x^{(n)}$,
and $\lambda$ is the coupling strength.

In line with (\ref{nmme_M}), we need to assign the correlation
function $c_{ab}(\tau)$. Here we use a bath correlation function
describing the electromagnetic field of a one-dimensional cavity \cite{Breuer2002},
\begin{eqnarray}\label{eq:bath_corr_split}
c^{(i)}(\tau)=& \frac{\pi}{2  \epsilon_0}\int^{\omega_c}_{0}
d\omega\,\omega
\left[ \cos(\omega \tau)\, \coth\left(\dfrac{\beta^{(i)}
\omega}{2}
\right) -i \sin (\omega \tau ) \right] ,
\end{eqnarray}
where $\beta^{(i)}$ is the inverse temperature of bath $i=L,R$.
Accordingly, one can calculate the power spectrum of this bath correlation
function as
\begin{equation}
\label{eq:bath_corr_final}
C^{(i)}(\omega) =  \frac{\pi^2|\omega|}{\epsilon_0}
  \left[ n_B(\beta^{(i)}|\omega|)
  + \theta(-\omega) \right]
  \quad \mbox{for}\quad |\omega|<\omega_{c}.
\end{equation}
This is the one-dimensional analogue of (\ref{cavity-power-spectrum}). 
One can immediately prove that this correlation function
does fulfil the detailed-balance relation and therefore we expect the system to
be driven towards thermal equilibrium if
the temperatures of the two baths are the same.

To investigate the energy transport, we identify the energy current according
to a continuity equation for the local energy. We define a
local Hamiltonian according to
\begin{equation}
\hat h^{(\mu)} = \frac{\Omega}{2}\, \sigma_z^{(\mu)}+\frac{\Gamma}{2}\,
  \big(\vec{\sigma}^{(\mu)}\cdot \vec{\sigma}^{(\mu+1)}
  + \vec{\sigma}^{(\mu-1)}\cdot
  \vec{\sigma}^{(\mu)}\big)
\end{equation}
if $\mu$ is different from $n$ and $1$.

We also define
\begin{equation}
\hat h^{(1)} = \frac{\Omega}{2}\,\sigma_z^{(1)}+\frac{\Gamma}{2}
\,\vec{\sigma}^{(1)}\cdot \vec{\sigma}^{(2)}
\end{equation}
and
\begin{equation}
\hat h^{(n)}=\frac{\Omega}{2}\,\sigma_z^{(n)}+\frac{\Gamma}{2}
\,\vec{\sigma}^{(n-1)}\cdot \vec{\sigma}^{(n)}
\end{equation}
so that $\hat H_S=\sum_\mu \hat h^{(\mu)}$. The time evolution of
this local Hamiltonian is given by
\begin{equation}
-\frac{d \hat h^{(\mu)}}{d t} = -i\, [\hat H_S,\hat h^{(\mu)}]
  =\hat j^{(\mu),(\mu+1)} - \hat j^{(\mu-1),(\mu)} ,
\label{energy-continuity}
\end{equation}
where energy-current operators have been defined as
\begin{eqnarray}
\hat j^{(\mu),(\mu+1)} = \frac{i}{4}\, \big[\Omega\,
  (\sigma_z^{(\mu)}-\sigma_z^{(\mu+1)}),\: \Gamma\, \vec{\sigma}^{(\mu)}\cdot
\vec{\sigma}^{(\mu+1)}\big].
\end{eqnarray}
Equation (\ref{energy-continuity}) has the form of a continuity equation for the
energy at site $\mu$ and is valid for sites
inside the spin chains that are not coupled to a bath. 

\begin{figure}[ht!]
\includegraphics[width=8cm]{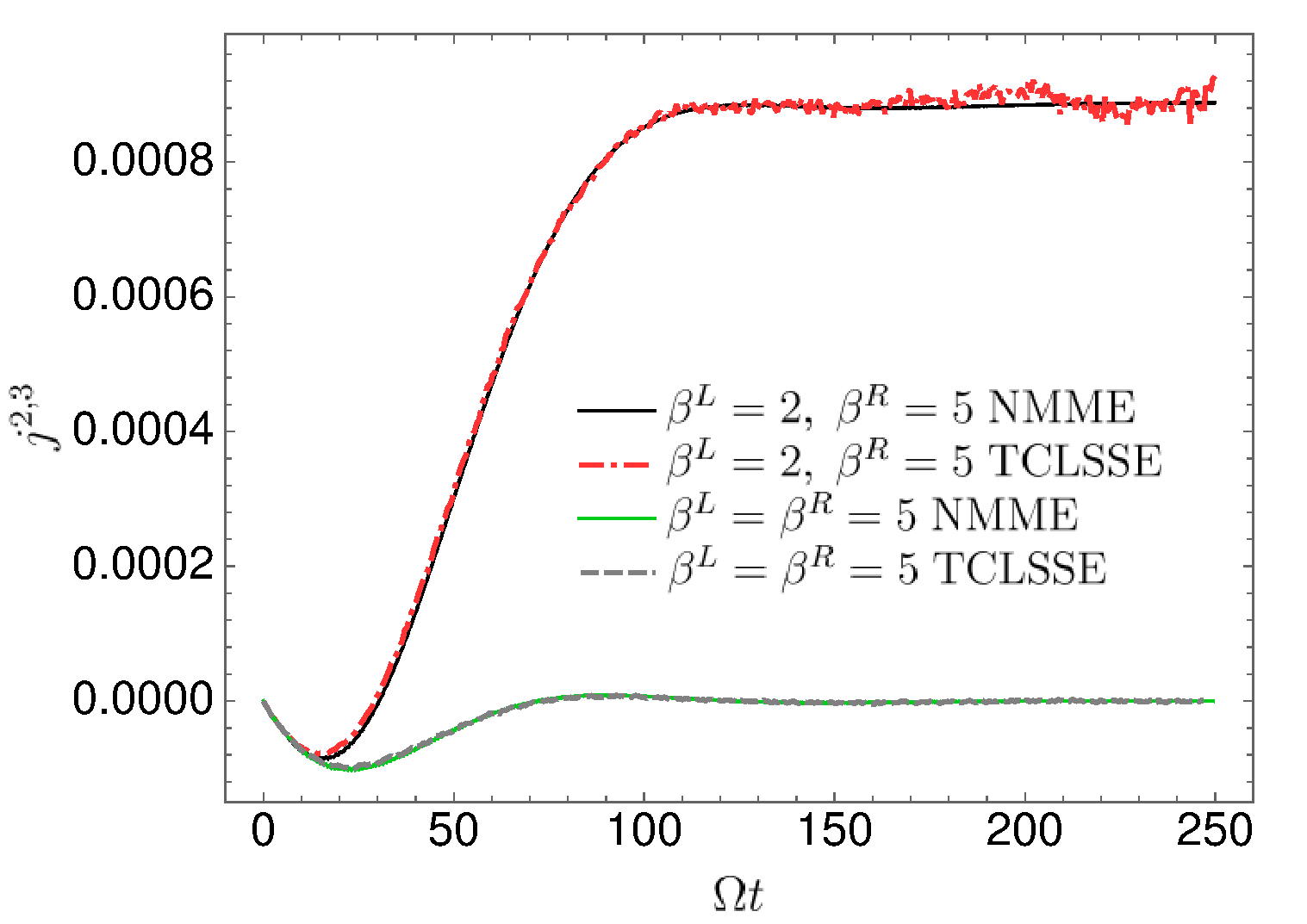}
\caption{Dynamics of the energy current of a three-site spin chain coupled
locally to two baths for the cases of equal and unequal
temperatures, calculated with the NMME (solid lines) and the TCLSSE
(dashed and dot-dashed lines). The agreement between the two sets
of lines is excellent, in particular at short times.
The time $t$ is measured in terms of the inverse of the energy
constant $\Omega$.}
\label{spin-chain}
\end{figure}

In figure \ref{spin-chain} we report the energy current flowing from the second
to the third spin of a three-site spin chain. In the equal-temperature case
($\beta^L=\beta^R=5$, green solid and grey dashed
lines in figure \ref{spin-chain}), a steady state is reached
for long times that coincides with the thermal equilibrium and
hence no current is flowing through the system.
On the other hand, for the case of unequal temperatures
($\beta^L=2$ and $\beta^R=5$), the
steady state shows a non-zero energy current from the warmer to the colder
bath, as expected (black solid and red dot-dashed
lines in figure \ref{spin-chain}).
For the TCLSSE we have averaged over $100\,000$ independent realisations of the
noise and for both calculations we have used the parameter values
$\Omega = 1$, $\Gamma = 0.01$, $\lambda = 0.1$, $\epsilon_0=1$
and $\omega_c=6$.
Both for the equal-temperature case and for the case with a thermal gradient,
we have chosen an initial state populated with the probabilities
determined by the equilibrium distribution at the lower
temperature. It can been seen that the TCLSSE produces the same dynamics of the energy current as obtained from the NMME in the equilibrium and non-equilibrium regime and hence can be seen as a reliable tool to simulate energy transport with moderate numerical cost.

\section{Conclusions}
In conclusion, we have numerically investigated a time-local
(time-convolutionless) version of a \emph{non-Markovian} stochastic
Schr\"odinger equation, which correctly describes the approach to thermal
equilibrium and energy transport as obtained from the general master
equation (\ref{nmme}). We report two case studies, which show that the TCLSSE is a
viable alternative for obtaining the exact dynamics of a non-Markovian open
quantum system. Moreover, contrary to other approximations, e.g., the
Born-Markov approximation to the Redfield equation \cite{Breuer2002}, this
stochastic equation reproduces the full dynamics of the non-Markovian master
equation, and therefore could be used to investigate the transient dynamics and
the approach to equilibrium. The TCLSSE can be integrated with moderate
numerical cost, comparable to that of a Markovian system. It also shows more
advantageous scaling with the number of states compared to the master equation,
which is particularly useful for large systems. We have also introduced an
efficient and portable numerical algorithm for the generation of the coloured
complex noise necessary to solve the time-convolutionless stochastic
Schr\"odinger equation. Our numerical algorithm is moderately faster than other
available algorithms and requires only the power spectrum, $C(\omega)$, of the
bath-correlation function as input.

\ack

R. B. and R. D'A. acknowledge support from MICINN (FIS2010-21282-C02-01 and
PIB2010US-00652), the Grupos Consolidados UPV/EHU del Gobierno Vasco
(IT-319-07) and ACI-Promociona (ACI2009-1036) and the financial support of the
CONSOLIDER-INGENIO 2010 ``NanoTherm'' (CSD2010-00044). R. B. acknowledges
financial support from IKERBASQUE, Basque Foundation for Science and the
Ministerio de Educaci\'on, Cultura y Deporte (FPU12/01576). C. T. acknowledges
financial support by the Deutsche Forschungsgemeinschaft, in part through
Research Unit FOR 1154 ``Towards Molecular Spintronics''. R.D'A. acknowledges
the support from the Diputacion Foral de Gipuzkoa via the grant number
Q4818001B and is thankful for its hospitality to the Physics Department of the
King's College London.

\appendix
\section{Derivation of the bath correlation function}

In this section, we give a derivation of the bath correlation function for
the coupling of the electromagnetic field in a three-dimensional
cavity of volume $V$ to an electronic system. In the dipole
approximation, the interaction Hamiltonian is
\begin{equation}
  \hat H_\mathrm{int} = - q \sum_i \hat {\vec{r}}_i \otimes \hat{\vec{E}}(t) ,
\end{equation}
where $q$ is the charge of an electron and $\hat{\vec{E}}$ is the electric field
inside the cavity. The wavelength of the electromagnetic field is assumed to be
large compared to the system size, hence $\hat{\vec{E}}$ is
considered to
be uniform in space. For simplicity, we suppose that each mode of the cavity has
the same polarization direction, $\vec u$. Thus the second-quantized form of
this interaction term is \cite{Schleich2001}
\begin{equation}
  \hat H_\mathrm{int} = -q\sum_{l,p}   \vec{u}\cdot \langle \psi_l | \hat{\vec r} | \psi_p 
    \rangle \hat \epsilon_l^\dagger\hat \epsilon_p \otimes 
      \sum_k i p_k \big( \hat b_k e^{-i \omega_k t} - \hat b_k^\dagger e^{i\omega_k t} \big)
	=\hat  S\otimes \hat B,
\label{interaction_cavity}
\end{equation}
where $\epsilon_l^\dagger$ creates an electron in the system in the state
$|\psi_l\rangle$. These states form an orthonormal basis of the
system Hamiltonian. The $k$-th field mode
inside the cavity with frequency $\omega_k$ is created by $\hat b^\dagger_k$ and
we define $p_k=\sqrt{ \omega_k /(2 V \epsilon_0)}$, where
$\epsilon_0$ is the dielectric constant.
We note, by comparing (\ref{interaction_cavity})
and (\ref{H_int}), that the coupling is already written in the required bilinear
form. From (\ref{interaction_cavity}) we can immediately read off
the form of the operators $\hat S$ and $\hat B$. After having assigned
these coupling operators, one can calculate the bath correlation
function
\begin{eqnarray}
\hspace*{-0.5cm}
c(t,\tau) 
&= \mbox{Tr}_B[\hat \rho_B^{\mathrm{eq}}\,\hat B(t)\,\hat B(\tau)]
 \nonumber \\
&=-\mbox{Tr}_B \left[ \hat \rho_B^{\mathrm{eq}} \sum_{k,j} p_k p_j \big(
  \hat b_k e^{-i \omega_k t} - \hat b_k^{\dagger}e^{i\omega_k t} \big)\big(\hat b_j e^{-i \omega_j \tau}- \hat b_j^{\dagger}e^{i \omega_j \tau}\big) \right].
\end{eqnarray}
Evaluating the trace in the bosonic many-particle basis of the bath and
replacing
the sum over the bath modes $\omega_k$ by an integral over 
frequency yields 
\begin{equation}
c(t,\tau) =\dfrac{1}{2  \epsilon_0 \pi^2}\int^{\omega_c}_{0} d\omega\,
  \omega^3 \bigg\{ \big[ n_B(\beta\omega)
  +1 \big]\, e^{-i\omega (t-\tau)} + 
  n_B(\beta\omega)\, e^{i \omega (t-\tau)}\bigg\},
\end{equation}
where we have inserted the density of states in the cavity, $\omega^2/\pi^2$.
We have introduced a cutoff frequency $\omega_c$ to be consistent with the
dipole approximation, which restricts the wavelengths of the bath
modes to be larger than the system size.
In addition, we note that the integral
cannot be evaluated analytically, whereas the power spectrum of
this function, (\ref{cavity-power-spectrum}),
is analytically known.


\section*{References}

\begin{thebibliography}{99}

\bibitem{Einstein1905}
Einstein A 1905 {\em Ann. Phys. (N. Y.)} {\bf 322} 549

\bibitem{Langevin1908}
Langevin P 1908 {\em C. R. Acad. Sci.} {\bf 146} 530

\bibitem{Kloeden1999}
Kloeden P~E and Platen E 1999 {\em Numerical Solution of Stochastic Differential Equations \/} (Heidelberg: Springer-Verlag)

\bibitem{Gardiner2000}
Gardiner C~W and Zoller P 2000 {\em Quantum Noise\/} 2nd ed
(Berlin: Springer)

\bibitem{Breuer2002}
Breuer H~P and Petruccione F 2002 {\em {The Theory of Open Quantum Systems}\/}
(New York: Oxford University Press)

\bibitem{Razavy2006}
Razavy M 2006 {\em {Classical and Quantum Dissipative Systems}\/} (London:
Imperial College Press)

\bibitem{vanKampen}
van Kampen N~G 2007 {\em {Stochastic Processes in Physics and Chemistry}\/} 3rd ed (Amsterdam: Elsevier)

\bibitem{Weiss2007}
Weiss U 2007 {\em {Quantum Dissipative Systems}\/} 3rd ed (Singapore: World Scientific)

\bibitem{Ghirardi1990}
Ghirardi G~C, Pearle P and Rimini A 1990 {\em Phys. Rev. A\/} {\bf 42} 78

\bibitem{Diosi1997}
Di\'{o}si L and Strunz W~T 1997 {\em Phys. Lett. A\/} {\bf 235} 569

\bibitem{Gaspard1999}
Gaspard P and Nagaoka M 1999 {\em J. Chem. Phys.\/} {\bf 111} 5676

\bibitem{Yu1999}
Yu T, Di\'{o}si L, Gisin N and Strunz W~T 1999 {\em Phys. Rev. A\/} {\bf 60} 91

\bibitem{Strunz2000}
Strunz W~T, Di\'{o}si L and Gisin N 2000 {\em Lect. Notes Phys.\/} {\bf 538} 271

\bibitem{DAgosta2008a}
D'Agosta R and {Di Ventra} M 2008 {\em Phys. Rev. B\/} {\bf 78} 165105

\bibitem{Pershin2008}
Pershin Y~V, Dubi Y and {Di Ventra} M 2008 {\em Phys. Rev. B\/} {\bf 78} 054302

\bibitem{Marques2006}
Marques M~A~L, Ullrich C~A, Rubio A, Nogueira F, Burke K and Gross E~K~U (eds) 2006 {\em {Time-Dependent Density Functional Theory}\/} ({\em Lecture Notes in Physics\/} vol 706) (Berlin: Springer)

\bibitem{Burke2005}
Burke K, Car R and Gebauer R 2005 {\em Phys. Rev. Lett.\/} {\bf 94} 146803

\bibitem{DiVentra2007}
{Di Ventra} M and D'Agosta R 2007 {\em Phys. Rev. Lett.\/} {\bf 98} 226403

\bibitem{Nak58}Nakajima S 1958 {\em Prog. Theor. Phys.\/} {\bf
20} 948

\bibitem{Zwa60}Zwanzig R 1960 {\em J. Chem. Phys.\/} {\bf 33}
1338; 1964 {\em Physica\/} {\bf 30} 1109

\bibitem{Wichterich2007}
Wichterich H, Henrich M, Breuer H~P, Gemmer J and Michel M 2007 {\em Phys. Rev. E\/} {\bf 76} 031115 

\bibitem{ToM76}Tokuyama M and Mori H 1976 {\em Prog. Theor.
Phys.\/} {\bf 55} 411

\bibitem{STH77}Hashitsume N, Shibata F and Shing\={u} M 1977 {\em
J. Stat. Phys.\/} {\bf 17} 155; Shibata F, Takahashi Y and Hashitsume N 1977
{\em J. Stat. Phys.\/} {\bf 17} 171

\bibitem{Timm2011}
Timm C 2011 {\em Phys. Rev. B\/} {\bf 83} 115416

\bibitem{Strunz2004}
Strunz W and Yu T 2004 {\em Phys. Rev. A\/} {\bf 69} 052115 

\bibitem{DeVega2005}
de~Vega I, Alonso D and Gaspard P 2005 {\em Phys. Rev. A\/} {\bf 71} 023812

\bibitem{DeVega2005b}
de~Vega I, Alonso D, Gaspard P and Strunz W~T 2005 {\em J. Chem. Phys.\/} {\bf 122} 124106 
  
\bibitem{Biele2011a}
Biele R and D'Agosta R 2012 {\em J. Phys. Condens. Matter\/} {\bf 24} 273201

\bibitem{Luczka2005}
\L uczka J 2005 {\em Chaos\/} {\bf 15} 26107 

\bibitem{Rice1944}
Rice S~O 1944 {\em Bell Syst. Tech. J.\/} {\bf 23} 2

\bibitem{Billah1990}
Billah K~Y~R and Shinozuka M 1990 {\em Phys. Rev. A\/} {\bf 42} 7492

\bibitem{Barrat2011}
Barrat J~L and Rodney D 2011 {\em J. Stat. Phys.\/} {\bf 144} 679

\bibitem{KGL10}Koller S, Grifoni M, Leijnse M and Wegewijs M~R 2010 {\em Phys.
Rev. B\/} {\bf 82} 235307

\bibitem{Mannella1992}
Mannella R and Palleschi V 1992 {\em Phys. Rev. A\/} {\bf 46} 8028

\bibitem{Press1992}
Press W~H, Teukolsky S~A, Vetterling W~T and Flannery B~P 2001 {\em Numerical
Recipes in Fortran 77: The Art of Scientific Computing} (Cambridge: Cambridge
University Press)

\bibitem{Kloeden1997}
Kloeden P~E, Platen E and Schurz H 1997 {\em Numerical Solution of SDE Through Computer Experiments \/} 3rd ed (Berlin: Springer-Verlag) 

\bibitem{Monasterio2007}
Mejia-Monasterio C and Wichterich, H 2007 {\em Eur. Phys. J. Spec. Top.} {\bf 151} 113

\bibitem{Schleich2001}
Schleich W~P 2001 {\em Quantum Optics in Phase Space \/} (Berlin:
Wiley-VCH)
   
\end{thebibliography}
\end{document}